\documentclass[aps,prl,reprint,floatfix,superscriptaddress,showpacs,citeautoscript]{revtex4-1}
\usepackage{graphicx}
\usepackage{bm,amsmath,amssymb,mathrsfs,dcolumn}
\usepackage{epsfig}
\usepackage{hyperref}
\usepackage[usenames]{color}
\hypersetup{pdfborder=0 0 0,colorlinks=true,citecolor=blue,linkcolor=blue}
\begin{document}
\title{Interface enhancement of Gilbert damping from first-principles}


\author{Yi Liu}
\altaffiliation[Present address: ]{Institut f{\"u}r Physik, Johannes Gutenberg--Universit{\"a}t Mainz, Staudingerweg 7, 55128 Mainz, Germany}
\affiliation{Faculty of Science and Technology and MESA$^+$ Institute for Nanotechnology, University of Twente, P.O. Box 217, 7500 AE Enschede, The Netherlands}
\author{Zhe Yuan}\email{zyuan@uni-mainz.de}
\affiliation{Faculty of Science and Technology and MESA$^+$ Institute for Nanotechnology, University of Twente, P.O. Box 217, 7500 AE Enschede, The Netherlands}
\affiliation{Institut f{\"u}r Physik, Johannes Gutenberg--Universit{\"a}t Mainz, Staudingerweg 7, 55128 Mainz, Germany}
\author{R. J. H. Wesselink}
\author{Anton A. Starikov}
\author{Paul J. Kelly}
\affiliation{Faculty of Science and Technology and MESA$^+$ Institute for Nanotechnology, University of Twente, P.O. Box 217, 7500 AE Enschede, The Netherlands}

\date{\today}

\begin{abstract}
The enhancement of Gilbert damping observed for Ni$_{80}$Fe$_{20}$ (Py) films in contact with the non-magnetic metals Cu, Pd, Ta and Pt, is quantitatively reproduced using first-principles scattering calculations.
The ``spin-pumping'' theory that qualitatively explains its dependence on the Py thickness is generalized to include a number of extra factors known to be important for spin transport through interfaces.
Determining the parameters in this theory from first-principles shows that interface spin-flipping makes an essential contribution to the damping enhancement. 
Without it, a much shorter spin-flip diffusion length for Pt would be needed than the value we calculate independently.
\end{abstract}

\maketitle

{\color{red}\it Introduction.}---Magnetization dissipation, expressed in terms of the Gilbert damping parameter $\alpha$, is a key factor determining the performance of magnetic materials in a host of applications. Of particular interest for magnetic memory devices based upon ultrathin magnetic layers \cite{[See the collection of articles ]HSTM,Bauer:natm12,Brataas:natm12} is the enhancement of the damping of ferromagnetic (FM) materials in contact with non-magnetic (NM) metals  \cite{Mizukami:jjap01,*Mizukami:jmmm01} that can pave the way to tailoring $\alpha$ for particular materials and applications. A ``spin pumping'' theory has been developed that describes this interface enhancement in terms of a transverse spin current generated by the magnetization dynamics that is pumped into and absorbed by the adjacent NM metal \cite{Tserkovnyak:prl02a,*Tserkovnyak:prb02b,Tserkovnyak:rmp05}. Spin pumping subsequently evolved into a technique to generate pure spin currents that is extensively applied in spintronics experiments \cite{Costache:prl06,Czeschka:prl11,Weiler:prl13}. 

\begin{figure}[b]
\begin{center}
\includegraphics[width=.95\columnwidth]{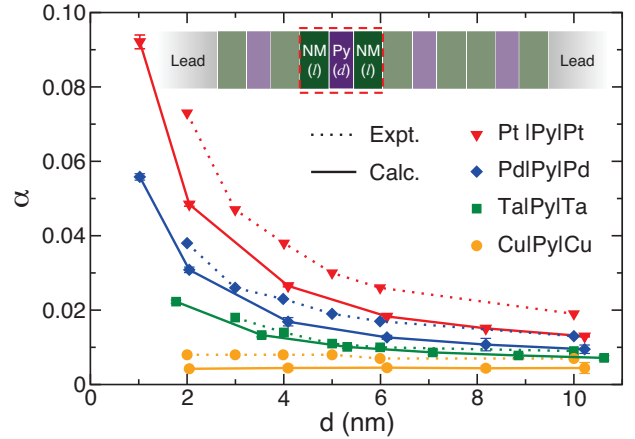}
\end{center}
\caption{(color online). Calculated (solid lines) Gilbert damping of NM$|$Py$|$NM (NM = Cu, Pd, Ta and Pt) compared to experimental measurements (dotted lines) \cite{Mizukami:jjap01} as a function of the Py thickness $d$. Inset: sketch of the structure used in the calculations. The dashed frame denotes one structural unit consisting of a Py film between two NM films. }\label{fig:1}
\end{figure}

A fundamental limitation of the spin-pumping theory is that it assumes spin conservation at interfaces. This limitation does not apply to a scattering theoretical formulation of the Gilbert damping that is based upon energy conservation, equating the energy lost by the spin system through damping to that parametrically pumped out of the scattering region by the precessing spins \cite{Brataas:prl08,*Brataas:prb11}.
In this Letter, we apply a fully relativistic density functional theory implementation \cite{Starikov:prl10,Liu:prb11,Yuan:arXiv14} of this scattering formalism to the Gilbert damping enhancement in those NM$|$Py$|$NM structures studied experimentally in Ref.~\onlinecite{Mizukami:jjap01,*Mizukami:jmmm01}. Our calculated values of $\alpha$ as a function of the Py thickness $d$ are compared to the experimental results in Fig.~\ref{fig:1}. Without introducing any adjustable parameters, we quantitatively reproduce the characteristic $1/d$ dependence as well as the dependence of the damping on the NM metal. 

To interpret the numerical results, we generalize the spin pumping theory to allow: (i) for interface \cite{Fert:prb96b,Eid:prb02,Bass:jpcm07} as well as bulk spin-flip scattering; (ii) the interface mixing conductance to be modified by spin-orbit coupling; (iii) the interface resistance to be spin-dependent. An important consequence of our analysis is that without interface spin-flip scattering, the value of the spin-flip diffusion length $l_{\rm sf}$ in Pt required to fit the numerical results is much shorter than a value we independently calculate for bulk Pt. A similar conclusion has recently been drawn for Co$|$Pt interfaces from a combination of ferromagnetic resonance, spin pumping and inverse spin Hall effect measurements \cite{Rojas-Sanchez:prl14}.  

{\color{red}\it Gilbert damping in NM$|$Py$|$NM.}---We focus on the NM$|$Py$|$NM sandwiches with NM = Cu, Pd, Ta and Pt that were measured in Ref.~\onlinecite{Mizukami:jjap01,*Mizukami:jmmm01}. The samples were grown on insulating glass substrates, the NM layer thickness was fixed at $l$=5~nm, and the Py thickness $d$ was variable. 
To model these experiments, the conventional NM-lead$|$Py$|$NM-lead two-terminal scattering geometry with semi-infinite ballistic leads \cite{Brataas:prl08,*Brataas:prb11,Starikov:prl10,Liu:prb11,Yuan:arXiv14} has to be modified because: (i) the experiments were carried out at room temperature so the 5 nm thick NM metals used in the samples were diffusive; (ii) the substrate$|$NM and NM$|$air interfaces cannot transmit charge or spin and behave effectively as ``mirrors'', whereas in the conventional scattering theory the NM leads are connected to charge and spin reservoirs.

We start with the structural NM$(l)|$Py$(d)|$NM$(l)$ unit indicated by the dashed line in the inset to Fig.~\ref{fig:1} that consists of a Py film, whose thickness $d$ is variable, sandwiched between $l$=5~nm-thick diffusive NM films. Several NM$|$Py$|$NM units are connected in series between semi-infinite leads to calculate the total magnetization dissipation of the system \cite{Brataas:prl08,*Brataas:prb11,Starikov:prl10,Liu:prb11,Yuan:arXiv14} thereby explicitly assuming a  ``mirror'' boundary condition. By varying the number of these units, the Gilbert damping for a single unit can be extracted 
\footnote{See Supplemental Material at http://link.aps.org/supplemental/10.1103/PhysRevLett.???.?????? for computational details.}, 
that corresponds to the damping measured for the experimental NM$(l)|$Py$(d)|$NM$(l)$ system. 

As shown in Fig.~\ref{fig:1}, the results are in remarkably good overall agreement with experiment. For Pt and Pd, where a strong damping enhancement is observed for thin Py layers, the values that we calculate are slightly lower than the measured ones. For Ta and Cu where the enhancement is weaker, the agreement is better. In the case of Cu, neither the experimental nor the calculated data shows any dependence on $d$ indicating a vanishingly small damping enhancement. The offset between the two horizontal lines results from a difference between the measured and calculated values of the bulk damping in Py. A careful analysis shows that the calculated values of $\alpha$ are inversely proportional to the Py thickness $d$ and approach the calculated bulk damping of Py $\alpha_0$=0.0046 \cite{Starikov:prl10} in the limit of large $d$ for all NM metals. However, extrapolation of the experimental data yields values of $\alpha_0$ ranging from 0.004 to 0.007 \cite{[See \S4.3.2 of ]Liu:14}; the spread can be partly attributed to the calibration of the Py thickness, especially when it is very thin.

{\color{red} \it Generalized spin-pumping theory.}---In spite of the very good agreement with experiment, our calculated results cannot be interpreted satisfactorily using the spin-pumping theory \cite{Tserkovnyak:prl02a,*Tserkovnyak:prb02b} that describes the damping enhancement in terms of a spin current pumped through the interface by the precessing magnetization giving rise to an accumulation of spins in the diffusive NM metal, and a back-flowing spin current driven by the ensuing spin-accumulation.
The pumped spin current, $\mathbf I_s^{\rm pump}=(\hbar^2 A/2e^2)G^{\rm mix}\mathbf m\times \dot{\mathbf m}$, is described using a ``mixing conductance'' $G^{\rm mix}$ \cite{Brataas:prl00} that is a property of the NM$|$FM interface \cite{Xia:prb02,Zwierzycki:prb05}. Here, $\mathbf m$ is a unit vector in the direction of the magnetization and $A$ is the cross-sectional area. The theory only takes spin-orbit coupling (SOC) into account implicitly via the spin-flip diffusion length $l_{\rm sf}$ of the NM metal and the pumped spin current is continuous across the FM$|$NM interface \cite{Tserkovnyak:prb02b}. 

With SOC included, this boundary condition does not hold. Spin-flip scattering at an interface is described by the ``spin memory loss'' parameter $\delta$ defined so that the spin-flip probability of a conduction electron crossing the interface is $1-e^{-\delta}$ \cite{Fert:prb96b,Eid:prb02}. It alters the spin accumulation in the NM metal and, in turn, the backflow into the FM material. To take $\delta$ and the spin-dependence of the interface resistance into account, the FM$|$NM interface is represented by a fictitious homogeneous ferromagnetic layer with a finite thickness \cite{Eid:prb02,Bass:jpcm07}. The spin current and spin-resolved chemical potentials (as well as their difference $\bm{\mu}_s$, the spin accumulation) are continuous at the boundaries of the effective ``interface'' layer. We impose the boundary condition that the spin current vanishes at NM$|$air or NM$|$substrate interfaces. Then the spin accumulation in the NM metal can be expressed as a function of the net spin-current $\mathbf I_s$ flowing out of Py 
\footnote{The derivation of spin accumulation is different from Ref.~\onlinecite{Rojas-Sanchez:prl14}, where Rojas-S{\'a}nchez et al. treated the interface region as a non-magnetic medium in their study of a Co$|$Pt interface. Since the resistance of a FM$|$NM interface is spin dependent \cite{Bass:jpcm07}, this is not strictly correct. In addition, their treatment contains an unknown empirical parameter, the ratio of the spin-conserved to spin-flip relaxation times.},
which is the difference between the pumped spin current $\mathbf I_s^{\rm pump}$ and the backflow $\mathbf I_s^{\rm back}$. The latter is determined by the spin accumulation in the NM metal close to the interface, $\mathbf I_s^{\rm back}[\bm{\mu}_s(\mathbf I_s)]$. Following the original treatment by Tserkovnyak et al. \cite{Tserkovnyak:prb02b}, $\mathbf I_s$ is determined by solving the equation $\mathbf I_s=\mathbf I_s^{\rm pump}-\mathbf I_s^{\rm back}[\bm{\mu}_s(\mathbf I_s)]$ self-consistently. Finally, the total damping of NM$(l)|$Py$(d)|$NM$(l)$ can be described as 
\begin{eqnarray}
&&\alpha(l,d) = \alpha_0 + \frac{g\mu_B\hbar}{e^2 M_s d} G_{\rm eff}^{\rm mix} 
              = \alpha_0 + \frac{g\mu_B\hbar}{e^2 M_s d}
\nonumber\\
&& \times\left[\frac{1}{G^{\rm mix}} +\frac{2\rho \, l_{\rm sf} \, R^{\ast}}{\rho \, l_{\rm sf} \, \delta \sinh{\delta}+R^{\ast}\cosh{\delta}\tanh(l/l_{\rm sf})}\right]^{-1}.
\label{eq:damping}
\end{eqnarray}
Here, $R^\ast=R/(1-\gamma^2_R)$ is an effective interface specific resistance with $R$ the total interface specific resistance between Py and NM and its spin polarization, $\gamma_R=(R^{\downarrow}-R^{\uparrow})/(R^{\downarrow}+R^{\uparrow})$ is determined by the contributions $R^{\uparrow}$ and $R^{\downarrow}$ from the two spin channels \cite{Bass:jpcm07}. $\rho$ is the resistivity of the NM metal. All the quantities in Eq.~(\ref{eq:damping}) can be experimentally measured \cite{Bass:jpcm07} and calculated from first-principles \cite{Brataas:prp06}. If spin-flip scattering at the interface is neglected, i.e., $\delta=0$, Eq.~(\ref{eq:damping}) reduces to the original spin pumping formalism \cite{Tserkovnyak:prb02b}. Eq.~(\ref{eq:damping}) is derived using the Valet-Fert diffusion equation \cite{Valet:prb93} that is still applicable when the mean free path is comparable to the spin-flip diffusion length \cite{Penn:prb05}.
 
\begin{figure}[t]
\begin{center}
\includegraphics[width=.9\columnwidth]{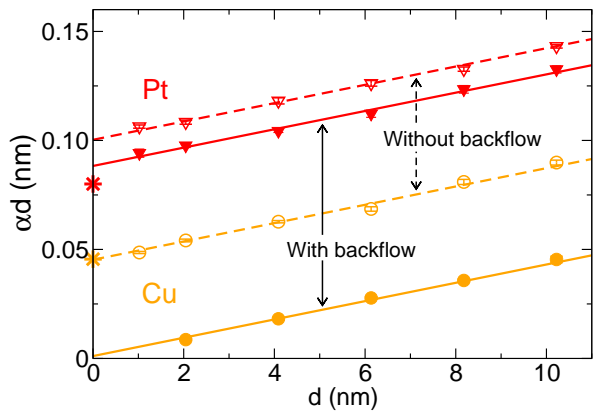}
\end{center}
\caption{(color online). Total damping calculated for Pt$|$Py$|$Pt and Cu$|$Py$|$Cu as a function of the Py thickness. The open symbols correspond to the case without backflow while the full symbols are the results shown in Fig.~\ref{fig:1} where backflow was  included. The lines are linear fits to the symbols. The asterisks on the $y$ axis are the values of $\overline{G}^{\rm mix}$ calculated without SOC using Eq.~(\ref{eq:mix}).}\label{fig:2}
\end{figure}

{\color{red} \it Mixing conductance.}---Assuming that SOC can be neglected and that the interface scattering is spin-conserving, the mixing conductance is defined as
\begin{equation}
\overline{G}^{\rm mix} =\frac{e^2}{h A}\sum_{m,n}\left(\delta_{mn}-r^{\uparrow}_{mn}r^{\downarrow\ast}_{mn}\right),\label{eq:mix}
\end{equation}
in terms of $r^{\sigma}_{mn}$, the probability amplitude for reflection of a NM metal state $n$ with spin $\sigma$ into a NM state $m$ with the same spin. Using Eq.~(\ref{eq:mix}), we calculate $\overline{G}^{\rm mix}$ for Py$|$Pt and Py$|$Cu interfaces without SOC and indicate the corresponding damping enhancement $g\mu_B\hbar \overline{G}^{\rm mix}/(e^2M_s A)$ on the vertical axis in Fig.~\ref{fig:2} with  asterisks. 

When SOC is included, Eq.~(\ref{eq:mix}) is no longer applicable. We can nevertheless identify a spin-pumping interface enhancement $G^{\rm mix}$ as follows. We artificially turn off the backflow by connecting the FM metal to ballistic NM leads so that any spin current pumped through the interface propagates away immediately and there is no spin accumulation in the NM metal. The Gilbert damping $\alpha d$ calculated without backflow (dashed lines) is linear in the Py thickness $d$; the intercept $\Gamma$ at $d=0$ represents an interface contribution. As seen in Fig.~\ref{fig:2} for Cu, $\Gamma$ coincides with the orange asterisk meaning that the interface damping enhancement for a Py$|$Cu interface is, within the accuracy of the calculation, unchanged by including SOC because this is so small for Cu, Ni and Fe. By contrast, $\Gamma$ and thus $G^{\rm mix}=e^2 M_s A\Gamma/(g\mu_B\hbar)$ for the Py$|$Pt interface is 25\% larger with SOC included, confirming the breakdown of Eq.~(\ref{eq:mix}) for interfaces involving heavy elements. 

\begin{table}[b]
\caption{Different mixing conductances calculated for Py$|$NM interfaces. $\overline{G}^{\rm mix}$ is calculated using Eq.~(\ref{eq:mix}) without SOC. $G^{\rm mix}$ is obtained from the intercept of the total damping $\alpha d$ calculated as a function of the Py thickness $d$ with SOC for ballistic NM leads. The effective mixing conductance $G^{\rm mix}_{\rm eff}$ is extracted from the effective $\alpha$ in Fig.~\ref{fig:1} in the presence of 5~nm NM on either side of Py. Sharvin conductances are listed for comparison. All values are given in units of $10^{15}~\Omega^{-1}\,\mathrm m^{-2}$.}
\begin{ruledtabular}
\begin{tabular}{lcccc}
NM  & $G_{\rm Sh}$ 
            & $\overline{G}^{\rm mix}$ 
                   & $G^{\rm mix}$ 
                          & $G^{\rm mix}_{\rm eff}$ \\
\hline
Cu  & 0.55  & 0.49 & 0.48 & 0.01 \\
Pd  & 1.21  & 0.89 & 0.98 & 0.57 \\
Ta  & 0.74  & 0.44 & 0.48 & 0.34 \\
Pt  & 1.00  & 0.86 & 1.07 & 0.95 \\
\end{tabular}
\end{ruledtabular}
\label{table:gmix}
\end{table}

The data in Fig.~\ref{fig:1} for NM=Pt and Cu are replotted as solid lines in Fig.~\ref{fig:2} for comparison. Their linearity means that we can extract an effective mixing conductance $G^{\rm mix}_{\rm eff}$ with backflow in the presence of 5~nm diffusive NM metal attached to Py. For Py$|$Pt, $G^{\rm mix}_{\rm eff}$ is only reduced slightly compared to $G^{\rm mix}$ because there is very little backflow. For Py$|$Cu, the spin current pumped into Cu is only about half that for Py$|$Pt. However, the spin-flipping in Cu is so weak that spin accumulation in Cu leads to a backflow that almost exactly cancels the pumped spin current and $G^{\rm mix}_{\rm eff}$ is vanishingly small for the Py$|$Cu system with thin, diffusive Cu. 

The values of $\overline{G}^{\rm mix}$, $G^{\rm mix}$ and $G^{\rm mix}_{\rm eff}$ calculated for all four NM metals are listed in Table~\ref{table:gmix}. Because $G^{\rm mix}$(Pd) and $G^{\rm mix}$(Pt) are comparable, Py pumps a similar spin current into each of these NM metals. The weaker spin-flipping and larger spin accumulation in Pd leads to a larger backflow and smaller damping enhancement. The relatively low damping enhancement in Ta$|$Py$|$Ta results from a small mixing conductance for the Ta$|$Py interface rather than from a large backflow. In fact, Ta behaves as a good spin sink due to its large SOC and the damping enhancement in Ta$|$Py$|$Ta can not be significantly increased by suppressing the backflow.

{\color{red}\it Thickness dependence of NM.}---In the following we focus on the Pt$|$Py$|$Pt system and examine the effect of changing the NM thickness $l$ on the damping enhancement, a procedure frequently used to experimentally determine the NM spin-flip diffusion length \cite{Azevedo:prb11,Liu:prl11,Kondou:ape12,Boone:jap13,Zhang:apl13}. 

\begin{figure}[b]
\begin{center}
\includegraphics[width=.9\columnwidth]{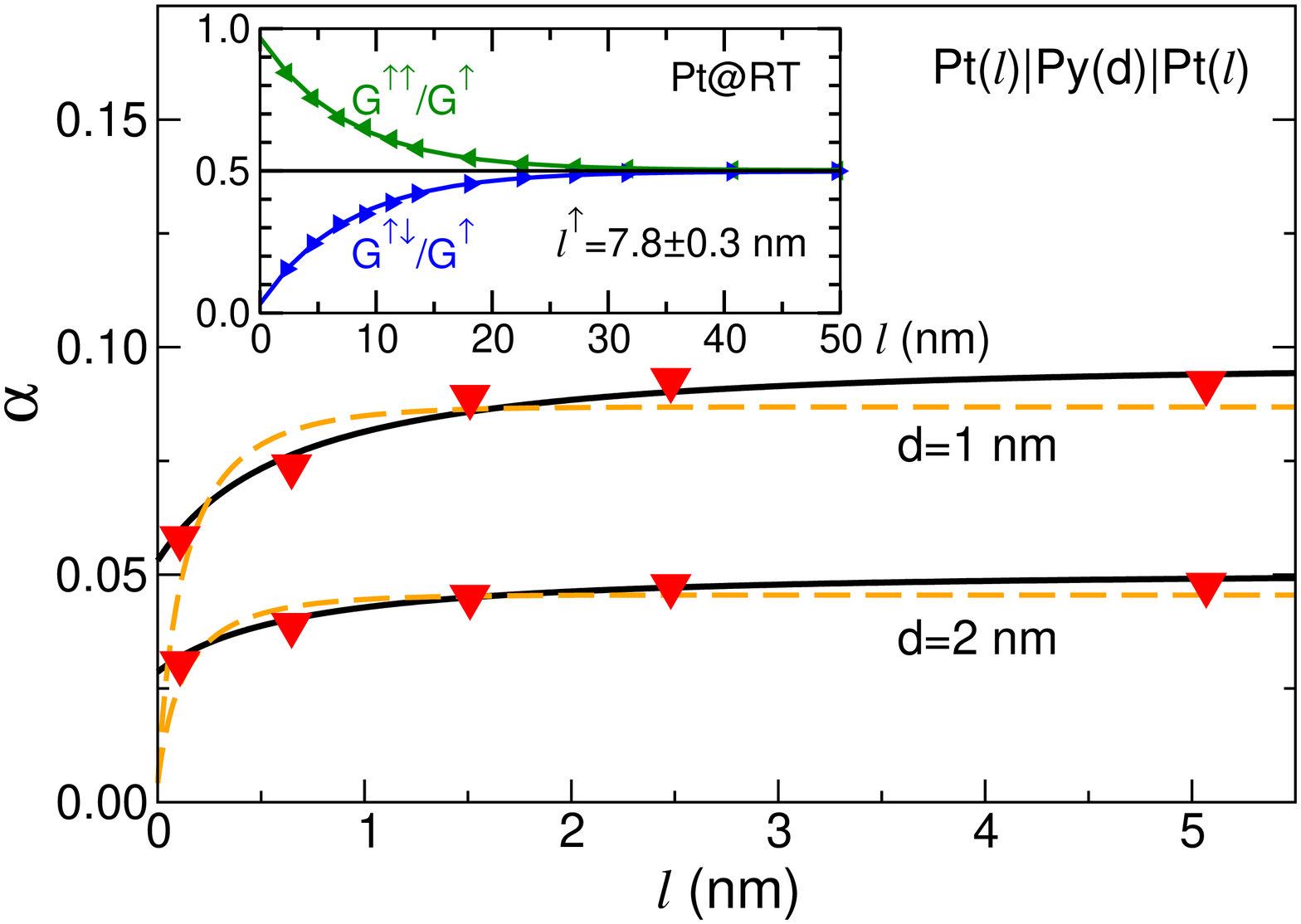}
\end{center}
\caption{$\alpha$ as a function of the Pt thickness $l$ calculated for Pt($l$)$|$Py($d$)$|$Pt($l$). The dashed and solid lines are the curves obtained by fitting without and with interface spin memory loss, respectively. Inset: fractional spin conductances $G^{\uparrow\uparrow}/G^{\uparrow}$ and $G^{\uparrow\downarrow}/G^{\uparrow}$ when a fully polarized up-spin current is injected into bulk Pt at room temperature. $G^{\sigma\sigma'}$ is ($e^2/h$ times) the transmission probability of a spin $\sigma$ from the left hand lead into a spin $\sigma'$ in the right hand lead; $G^{\uparrow}=G^{\uparrow\uparrow}+G^{\uparrow\downarrow}$. The value of the spin-flip diffusion length for a single spin channel obtained by fitting is $l^{\sigma}=7.8\pm0.3$~nm.}\label{fig:3}
\end{figure}

The total damping calculated for Pt$|$Py$|$Pt is plotted in Fig.~\ref{fig:3} as a function of the Pt thickness $l$ for two thicknesses $d$ of Py. For both $d=1$~nm and $d=2$~nm, $\alpha$ saturates at $l$=1--2~nm in agreement with experiment \cite{Liu:prl11,Kondou:ape12,Zhang:apl13,Boone:jap13,Rojas-Sanchez:prl14}. A fit of the calculated data using Eq.~(\ref{eq:damping}) with $\delta\equiv0$ requires just three parameters, $G^{\rm mix}$, $\rho$ and $l_{\rm sf}$. A separate calculation gives $\rho=10.4~\mu\Omega\,\mathrm{cm}$ at T=300~K in very good agreement with the experimental bulk value of $10.8~\mu\Omega\,\mathrm{cm}$ \cite{[See \S12: Properties of Solids - Electrical Resistivity of Pure Metals in ]HCP84}. Using the calculated $G^{\rm mix}$ from Table~\ref{table:gmix} leaves just one parameter free; from fitting, we obtain a value $l_{\rm sf}$=0.8~nm for Pt (dashed lines) that is consistent with values between 0.5 and 1.4~nm determined from spin-pumping experiments \cite{Liu:prl11,Kondou:ape12,Zhang:apl13,Boone:jap13}. However, the dashed lines clearly do not reproduce the calculated data very well and the fit value of $l_{\rm sf}$ is much shorter than that extracted from scattering calculations \cite{Starikov:prl10}. By injecting a fully spin-polarized current into diffusive Pt, we find $l^{\uparrow}=l^{\downarrow}=7.8\pm0.3$~nm, as shown in the inset to Fig.~\ref{fig:3}, and from \cite{Valet:prb93,Costache:13}, $l_{\rm sf}=\left[(l^{\uparrow})^{-2}+(l^{\downarrow})^{-2}\right]^{-1/2}=5.5\pm0.2$~nm. This value is confirmed by examining how the current polarization in Pt is distributed locally \cite{Wesselink:unpublished14}. 

If we allow for a finite value of $\delta$ and use the independently determined $G^{\rm mix}$, $\rho$ and $l_{\rm sf}$, the data in Fig.~\ref{fig:3} (solid lines) can be fit with $\delta=3.7\pm0.2$ and $R^\ast/\delta=9.2\pm1.7~\mathrm f\Omega\,\mathrm{m}^2$. The solid lines reproduce the calculated data much better than when $\delta=0$ underlining the importance of including interface spin-flip scattering \cite{Rojas-Sanchez:prl14,Nguyen:jmmm14}. The large value of $\delta$ we find is consistent with a low spin accumulation in Pt and the corresponding very weak backflow at the Py$|$Pt interface seen in Fig.~\ref{fig:2}.

\begin{figure}[t]
\begin{center}
\includegraphics[width=.9\columnwidth]{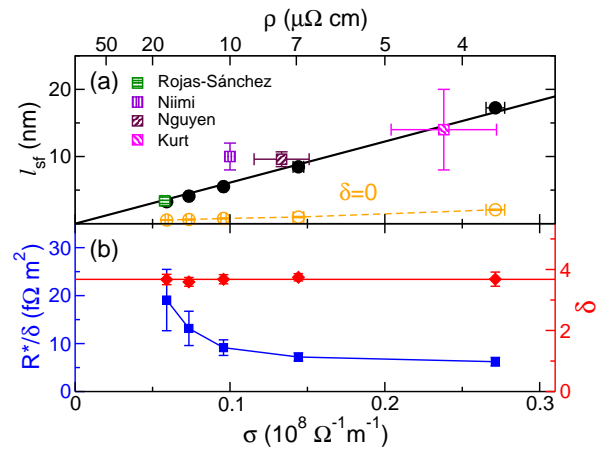}
\end{center}
\caption{(a) $l_{\rm sf}$ for diffusive Pt as a function of its conductivity $\sigma$ (solid black circles) calculated by injecting a fully polarized current into Pt. The solid black line illustrates the linear dependence. Bulk values extracted from experiment that are either not sensitive to interface spin-flipping \cite{Niimi:prl13} or take it into account \cite{Kurt:apl02,Rojas-Sanchez:prl14,Nguyen:jmmm14}
are plotted (squares) for comparison. The empty circles are values of $l_{\rm sf}$ determined from the interface-enhanced damping using Eq.~(\ref{eq:damping}) with $\delta=0$. (b) Fit values of $R^\ast/\delta$ and $\delta$ as a function of the conductivity of Pt obtained using Eq.~(\ref{eq:damping}). The solid red line is the average value (for different values of $\sigma$) of $\delta$=3.7.}
\label{fig:4}
\end{figure}

{\color{red}\it Conductivity dependence.}---Many experiments determining the spin-flip diffusion length of Pt have reported Pt resistivities that range from 4.2--12~$\mu\Omega\,\mathrm{cm}$ at low temperature \cite{Morota:prb11,Niimi:prl13,Kurt:apl02,Nguyen:jmmm14} and 15--73~$\mu\Omega\,\mathrm{cm}$ at room temperature \cite{Althammer:prb13,Castel:apl12,Ando:prl08,Rojas-Sanchez:prl14}. The large spread in resistivity can be attributed to different amounts of structural disorder arising during fabrication, the finite thickness of thin film samples etc. We can determine $l_{\rm sf}$ and $\rho \equiv 1/\sigma$ from first principles scattering theory \cite{Starikov:prl10,Liu:prb11} by varying the temperature in the thermal distribution of Pt displacements in the range 100--500~K. The results are plotted (black solid circles) in Fig.~\ref{fig:4}(a). $l_{\rm sf}$ shows a linear dependence on the conductivity suggesting that the Elliott-Yafet mechanism~\cite{Elliott:pr54,Yafet:63} dominates the conduction electron spin relaxation. 
A linear least squares fit yields $\rho \, l_{\rm sf}=0.61 \pm 0.02 \, {\rm f} \Omega \, {\rm m^2}$ that agrees very well with bulk data extracted from experiment that are either not sensitive to interface spin-flipping \cite{Niimi:prl13} or take it into account \cite{Kurt:apl02,Rojas-Sanchez:prl14,Nguyen:jmmm14}. For comparison, we plot values of $l_{\rm sf}$ extracted from the interface-enhanced damping calculations assuming
$\delta=0$ (empty orange circles). The resulting values of $l_{\rm sf}$ are very small, between 0.5 and 2~nm, to compensate for the neglect of $\delta$. 

Having determined $l_{\rm sf}(\sigma)$, we can calculate the interface-enhanced damping for Pt$|$Py$|$Pt for different values of $\sigma_{\rm Pt}$ and repeat the fitting of Fig.~\ref{fig:3} using Eq.~(\ref{eq:damping}) \footnote{Experiment \cite{Czeschka:prl11} and theory \cite{Nakata:jpcm13} indicate at most a weak temperature dependence of the spin mixing conductance.}. The parameters $R^\ast/\delta$ and $\delta$ are plotted as a function of the Pt conductivity in Fig.~\ref{fig:4}(b). The spin memory loss $\delta$ does not show any significant variation about 3.7, i.e., it does not appear to depend on temperature-induced disorder in Pt indicating that it results mainly from scattering of the conduction electrons at the abrupt potential change of the interface. Unlike $\delta$, the effective interface resistance $R^\ast$ decreases with decreasing disorder in Pt and tends to saturate for sufficiently ordered Pt. It suggests that although lattice disorder at the interface does not dissipate spin angular momentum, it still contributes to the relaxation of the momentum of conduction electrons at the interface. 

{\color{red}\it Conclusions.}---We have calculated the Gilbert damping for Py$|$NM-metal interfaces from first-principles and reproduced quantitatively the experimentally observed damping enhancement. To interpret the numerical results, we generalized the spin-pumping expression for the damping to allow for interface spin-flipping, a mixing conductance modified by SOC, and spin dependent interface resistances. The resulting Eq.~(\ref{eq:damping}) allows the two main factors contributing to the interface-enhanced damping to be separated: the mixing conductance that determines the spin current pumped by a precessing magnetization and the spin accumulation in the NM metal that induces a backflow of spin current into Py and lowers the efficiency of the spin pumping. In particular, the latter is responsible for the low damping enhancement for Py$|$Cu while the weak enhancement for Py$|$Ta arises from the small mixing conductance. 

We calculate how the spin-flip diffusion length, spin memory loss and interface resistance depend on the conductivity of Pt.
It is shown to be essential to take account of spin memory loss to extract reasonable spin-flip diffusion lengths from interface damping. 
This has important consequences for using spin-pumping-related experiments to determine the Spin Hall angles that characterize the Spin Hall Effect \cite{Rojas-Sanchez:prl14}.

{\color{red}\it Acknowledgments.}---We are grateful to G.E.W. Bauer for a critical reading of the manuscript. Our work was financially supported by the ``Nederlandse Organisatie voor Wetenschappelijk Onderzoek'' (NWO) through the research programme of ``Stichting voor Fundamenteel Onderzoek der Materie'' (FOM) and the  supercomputer facilities of NWO ``Exacte Wetenschappen (Physical Sciences)''. It was also partly supported by the Royal Netherlands Academy of Arts and Sciences (KNAW). Z. Yuan acknowledges the financial support of the Alexander von Humboldt foundation.


%

\end{document}